\documentclass[10pt]{article}
\usepackage{graphicx}

\def\Title#1{\begin{center} {\Large #1 } \end{center}}
\def\Author#1{\begin{center}{ \sc #1} \end{center}}
\def\Address#1{\begin{center}{ \it #1} \end{center}}

\newcommand\pubblock{\rightline{\begin{tabular}{l} Proceedings of the Second Annual LHCP\\ \pubnumber\\
         \pubdate  \end{tabular}}}

\newenvironment{Abstract}{\begin{quotation} \begin{center} 
             \large ABSTRACT \end{center}\bigskip 
      \begin{center}\begin{large}}{\end{large}\end{center} \end{quotation}}

\newenvironment{Presented}{\begin{quotation} \begin{center} 
             PRESENTED AT\end{center}\bigskip 
      \begin{center}\begin{large}}{\end{large}\end{center} \end{quotation}}

\def\Acknowledgements{\bigskip  \bigskip \begin{center} \begin{large}
             \bf ACKNOWLEDGEMENTS \end{large}\end{center}}
\usepackage{amssymb}
\usepackage{amsmath}
\usepackage{amsfonts}
\usepackage{graphicx}
\usepackage{color}
\usepackage{xspace}
\usepackage{ulem}

\catcode`\@=11
\font\manfnt=manfnt
\def\Watchout{\@ifnextchar [{\W@tchout}{\W@tchout[1]}}
\def\W@tchout[#1]{{\manfnt\@tempcnta#1\relax%
  \@whilenum\@tempcnta>\z@\do{%
    \char"7F\hskip 0.3em\advance\@tempcnta\m@ne}}}
\let\foo\W@tchout
\def\dubious{\@ifnextchar[{\@dubious}{\@dubious[1]}}

\def\@dubious[#1]{%
  \color{red}\setbox\@tempboxa\hbox{\@W@tchout#1}
  \@tempdima\wd\@tempboxa
  \list{}{\leftmargin\@tempdima}\item[\hbox to 0pt{\hss\@W@tchout#1}]}
\def\@W@tchout#1{\W@tchout[#1]}
\catcode`\@=12

\newcommand\pubnumber{ }

\newcommand\pubdate{}

\def\affiliation{
Department of Physics and Astronomy, University College London,\\
London WC1E 6BT, United Kingdom}

\begin{document}

% large size for the first page
\large
\begin{titlepage}
\pubblock

%% Change the title, name, abstract
%% Title 
\vfill
\Title{Impact of Lepton Number Violation at the LHC on Models of Leptogenesis}
\vfill

\Author{Frank F. Deppisch and Julia Harz}
\Address{\affiliation}
\vfill

\begin{Abstract}
\noindent 
The discovery of lepton number violation (LNV) at the LHC would have profound consequences for the viability of high-scale leptogenesis models. As an example, we discuss the case of observing a signal with two same-sign leptons, two jets and no missing energy. This would imply a large washout factor for the lepton number density in the early Universe, which leads to a significant constraint on any high-scale model for the generation of the observed baryon asymmetry. In a standard leptogenesis scenario, the corresponding washout factor would strongly decrease a pre-existing lepton asymmetry and thus would render leptogenesis models that generate a $(B-L)$ asymmetry far above the LHC scale ineffective. Therefore, LHC searches focused on LNV processes without missing energy are powerful probes for high-scale leptogenesis models and correspondingly shed light on the nature of baryogenesis and neutrino masses.

\end{Abstract}
\vfill

% DO NOT CHANGE 
\begin{Presented}
The Second Annual Conference\\
 on Large Hadron Collider Physics \\
Columbia University, New York, U.S.A \\ 
June 2-7, 2014
\end{Presented}
\vfill
\end{titlepage}
\def\thefootnote{\fnsymbol{footnote}}
\setcounter{footnote}{0}
%

% normal size for the rest
\normalsize 

%% Your paper should be entered below. 
\section{Introduction}
The observed baryon asymmetry of the Universe, quantified in terms of the baryon-to-photon number density ratio~\cite{Ade:2013zuv}
\begin{equation}
\label{eq:etaBobs}
  \eta_B^\mathrm{obs} = \left(6.20 \pm 0.15\right) \times 10^{-10},
\end{equation}
is far too large compared to the Standard Model (SM) expectation. This is because in SM electroweak (EW) baryogenesis, the necessary $CP$ violation is too small and no first order phase transition can take place due to the too large Higgs mass~\cite{Gavela}. The mechanism of baryogenesis in the Universe thus remains unexplained and beyond the SM physics has to be evoked. A large number of possible mechanisms and scenarios to generate the observed baryon asymmetry have been proposed in the literature. Due to the presence of non-perturbative $(B+L)$-violating sphaleron processes at and above the EW breaking scale of the SM~\cite{KRS}, many models work by generating an asymmetry in the quantum number $(B-L)$ above the EW scale, which then results in a baryon asymmetry after the sphaleron processes fall out of equilibrium. This involves a mechanism satisfying the three Sakharov conditions of non-equilibrium, $C+CP$ violation and $(B-L)$ violation (B violation is then provided by the sphalerons)~\cite{Sakharov}. As the presence of $(B-L)$ violation is a crucial ingredient, such scenarios can only work if the corresponding interactions are sufficiently out of equilibrium between the scale of the $(B-L)$ asymmetry generation and the EW scale, otherwise the asymmetry is washed out before the sphalerons take effect. 

One of the most popular realizations of this idea is through the mechanism of leptogenesis~\cite{FY}. In the standard leptogenesis scenario, the lepton asymmetry is generated by the out-of-equilibrium decay of very heavy, right-handed neutrinos, whereas their inverse decays and other $\Delta L = 1,2$ processes washout the asymmetry. Leptogenesis is very appealing as it incorporates the Seesaw mechanism of neutrino mass generation and the necessary lepton number violation (LNV) is tightly connected to the Majorana nature of the heavy and light neutrinos. An important downside of high-scale leptogenesis models is that they are difficult to test as their scale $\Lambda \gtrsim 10^{10}$~GeV is too far removed to be probed in laboratory experiments. It is an interesting question whether it is possible to test the mechanism of leptogenesis or if one is at least able to falsify leptogenesis. We address this issue by looking at LNV processes at the LHC. Especially, combining cosmology with collider physics provides a powerful possibility to further explore this question. The issue of falsifying leptogenesis at the LHC has been studied in model-specific contexts, see e.g.~\cite{Frere:2008ct}. Our analysis, however, as published in Ref.~\cite{Deppisch:2013jxa} and on which this proceedings report is based, focuses instead on a model-independent approach.

In the following, we discuss how the observation of lepton number violating processes with $\Delta L \neq 0$ and $\Delta (B-L) \neq 0$ without missing energy at the LHC gives rise to a stringent lower bound on the corresponding washout rate in the early Universe and leads to an exclusion of high-scale leptogenesis models. It has to be stressed that the non-observation of LNV does not allow to falsify leptogenesis.

\section{Lepton Number Violation at the LHC}
\begin{figure}[t!]
\centering
\vskip-6mm
\includegraphics[clip,width=0.3\linewidth]{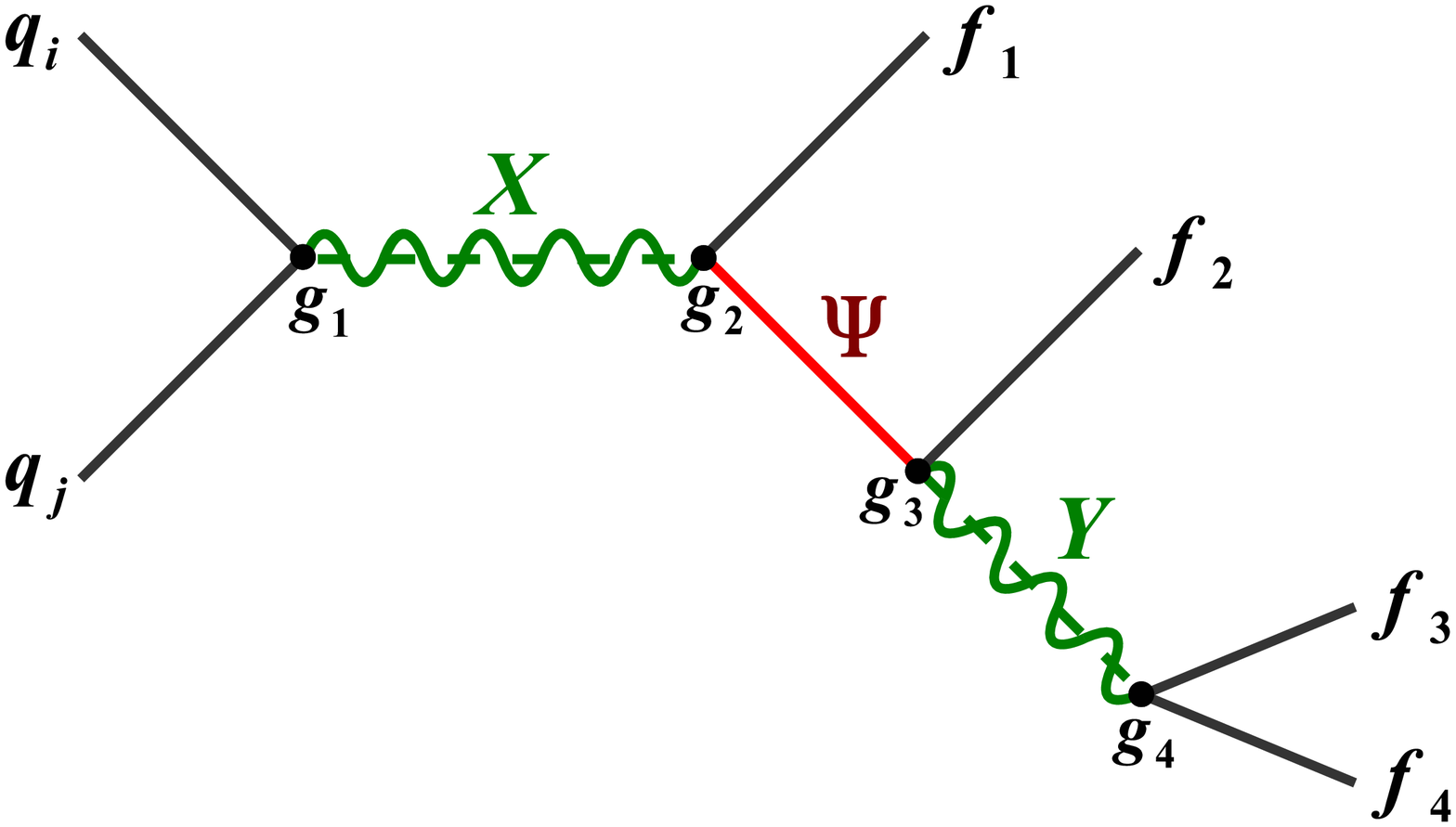}
\includegraphics[clip,width=0.3\linewidth]{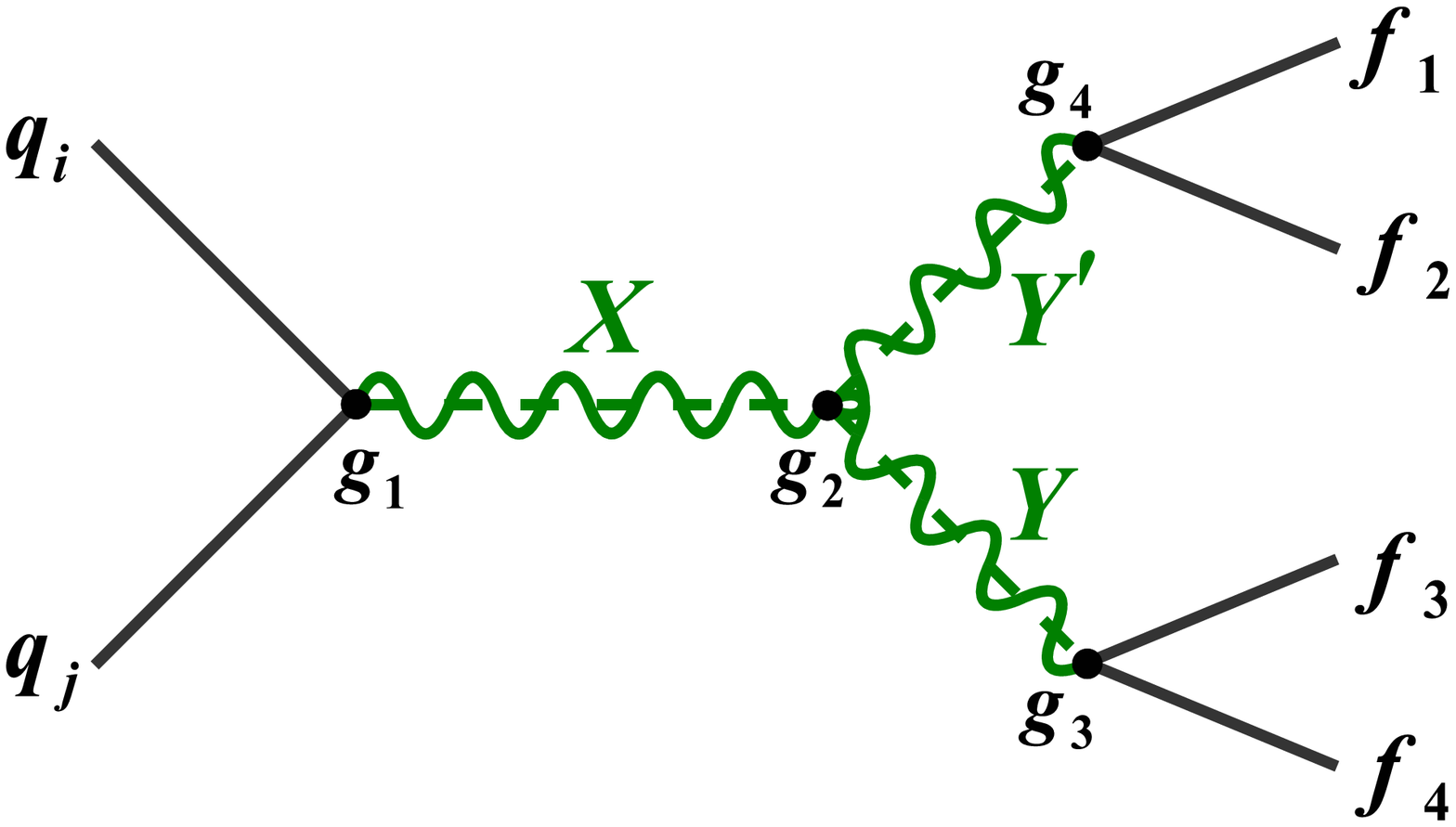}
\vskip-5mm
\caption{Diagrams producing a resonant same sign dilepton signal $pp \to l^\pm l^\pm q q$ at the LHC. The intermediate particles are generic scalar or vector bosons $X$, $Y^{(')}$ and a fermion $\Psi$ which interact via unspecified couplings with strengths $g_i$. Generally, any two of the out-going fermions $f_i$ can be leptons.}
\label{fig:decompositions}  
\end{figure}

Out of the possible LNV processes at the LHC, the analysis \cite{Deppisch:2013jxa} focuses on the resonant process $pp \to l^\pm l^\pm q q$ with two same-sign leptons and two jets without missing energy. The general tree-level diagrams for this process involve generic intermediate particles as shown in Fig.~\ref{fig:decompositions}. Depending on the SM quantum numbers of the intermediate particles, the initial state can consist of any combination of quark pairs. Likewise, any two of the four out-going fermions can be leptons~\cite{Helo:2013ika}. The requirement of a process without missing energy is crucial as the presence of LNV cannot be established otherwise. An example in this class of processes is resonant $W_R$ production in left-right symmetric models~\cite{Keung:1983uu}. 

The total LHC cross section of any resonant process described by Fig.~\ref{fig:decompositions} can be written as~\cite{Deppisch:2013jxa}
\begin{align}
	\label{eq:cs_lhc}
	\sigma_\text{LHC} =
	\frac{4\pi^2}{9 s}(2J_X+1)
	\frac{\Gamma_X}{M_X}
	f_{q_1 q_2}\!\left(\frac{M_X}{\sqrt{s}} , M_X^2 \right) 
	\times \text{Br}(X\to q_1 q_2)\times\text{Br}(X\to 4f),
\end{align}
within the narrow-width approximation of the resonance $X$. Here, $J_X$ is the spin of $X$, and $\text{Br}(X\to q_1 q_2)$ and $\text{Br}(X\to 4f)$ are its decay branching ratios into the initial state partons $q_i$ and the final state fermions, respectively. The LHC centre-of-mass energy is $\sqrt{s}$ for which we use 14~TeV throughout. The function $f_{q_1 q_2}\!\left(M_X /\sqrt{s}, M_X^2 \right)$ arises from integrating the quark level cross section over the relevant parton distributions. For masses $M_X\approx 1-5$~TeV, it can be well approximated as $f_{q_1 q_2}(M_X / \sqrt{s}) \approx A_{q_1 q_2} \times \exp(- C_{q_1 q_2} M_X / \sqrt{s})$~\cite{Leike:1998wr}. The coefficients range from $A_{\bar u\bar u} \approx 200$ to $A_{uu} \approx 4400$ and $C_{uu} \approx 26$ to $C_{\bar d\bar d} \approx 51$.

\begin{figure}[t]
\centering
\includegraphics[clip,width=0.24\linewidth]{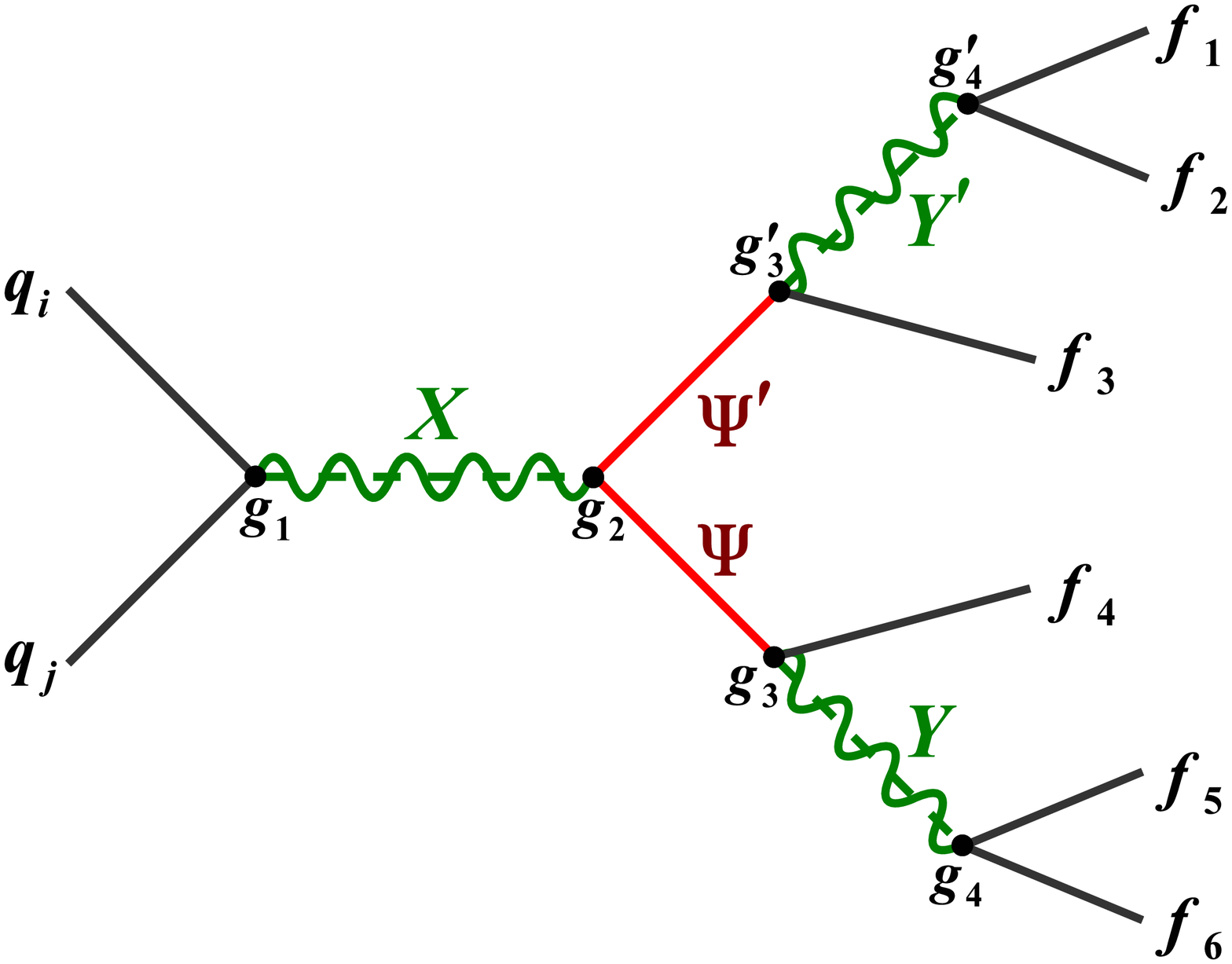}
\includegraphics[clip,width=0.24\linewidth]{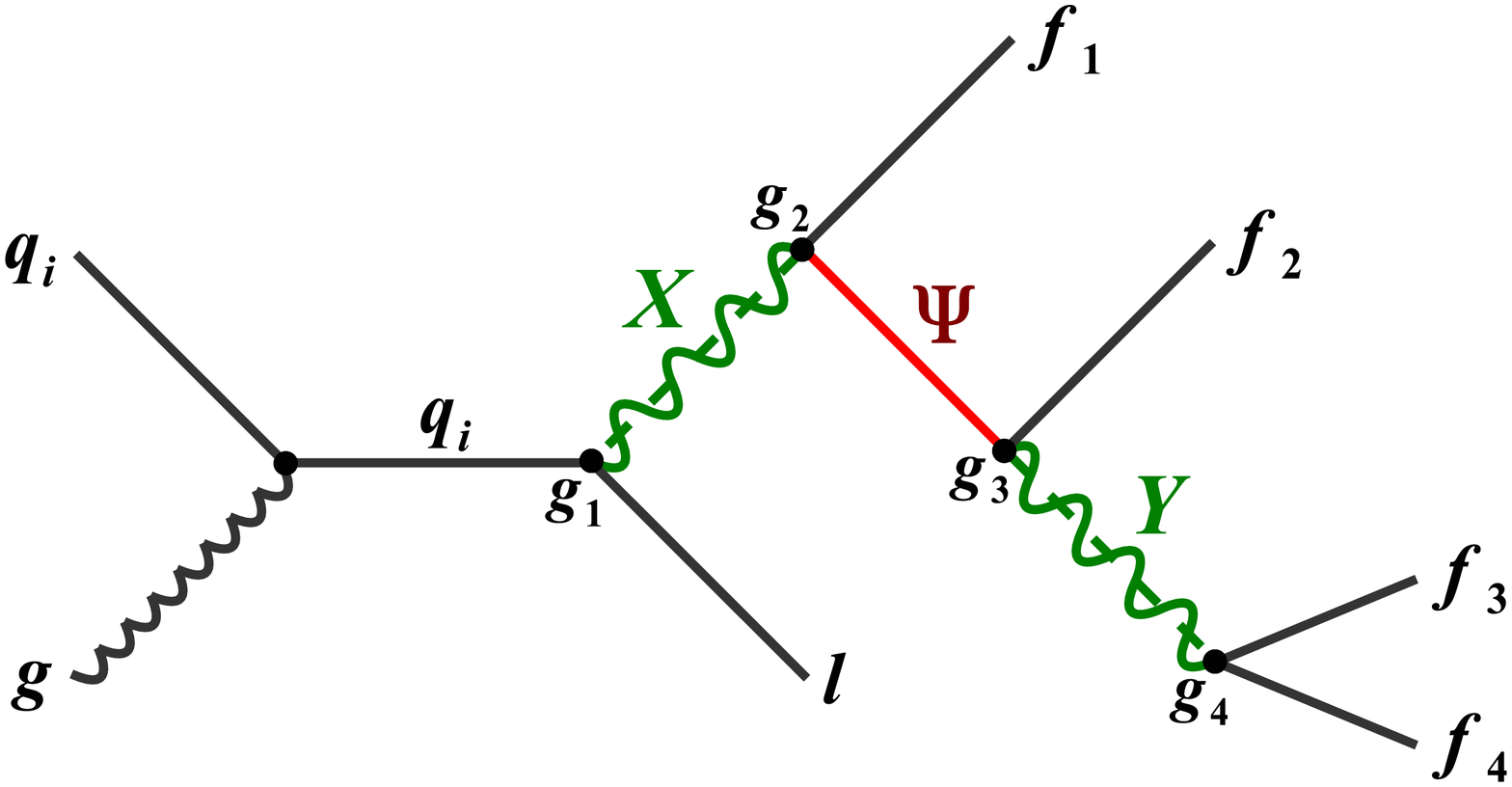}
\includegraphics[clip,width=0.24\linewidth]{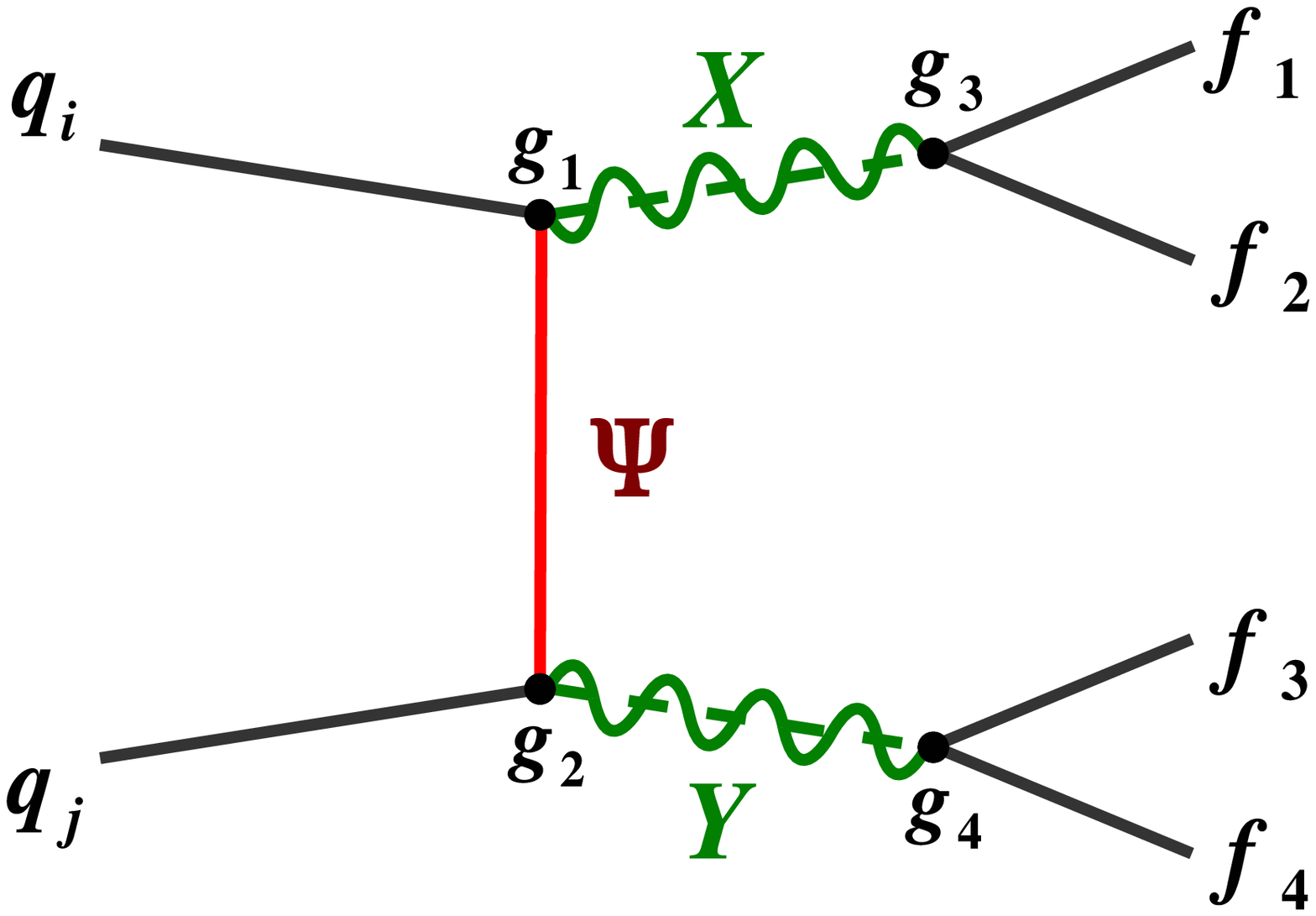}
\includegraphics[clip,width=0.24\linewidth]{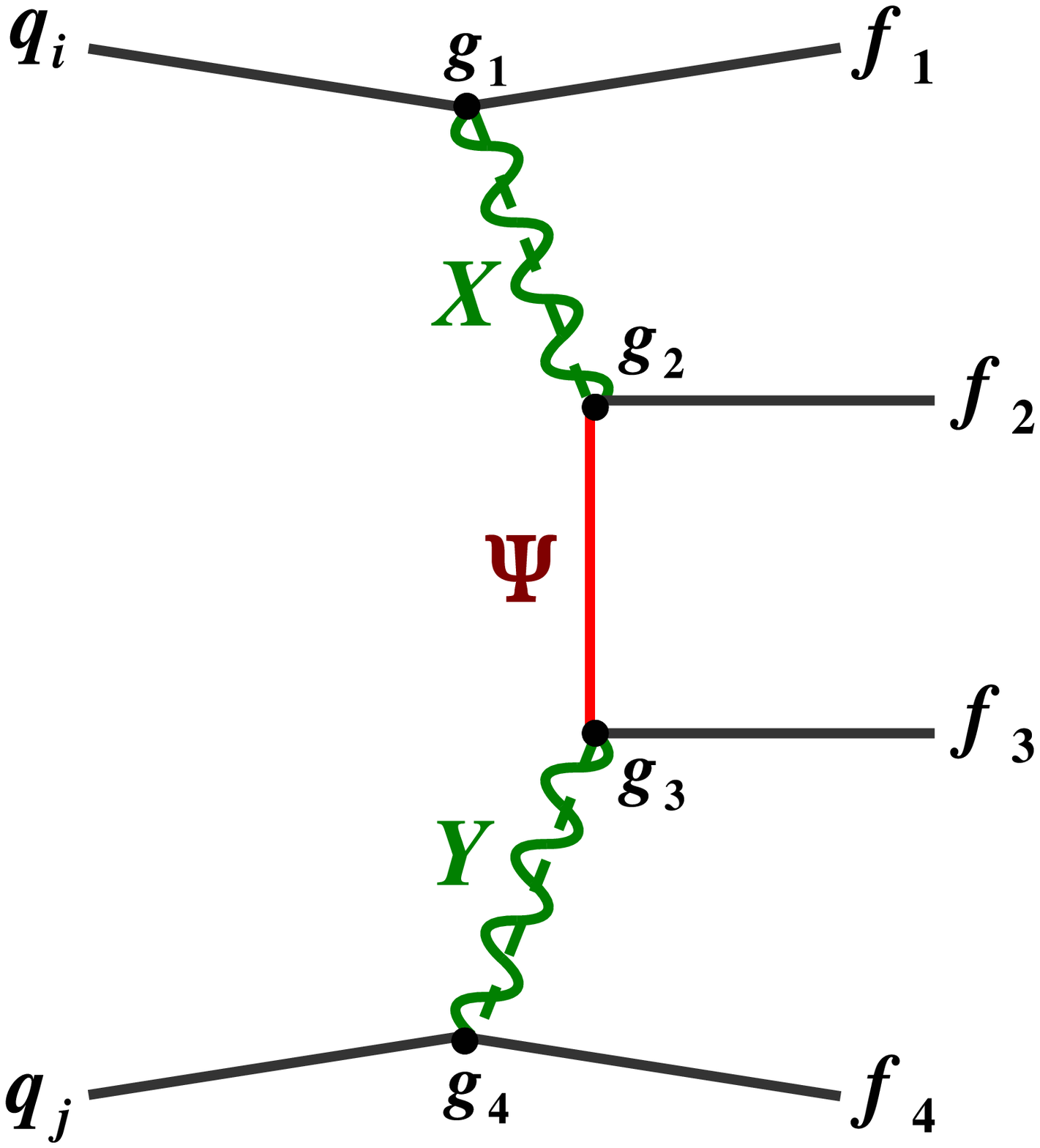}
\caption{Examples of other LNV processes at the LHC (left to right): (i) Resonant pair-production with six final state fermions; (ii) Associated production of leptoquark-like particle $X$ with five final state fermions; (iii, iv) Non-resonant processes with four final state fermions.}
\label{fig:decompositions2}  
\end{figure}

While the resonant process discussed so far simplifies the analysis and makes the connection to the cosmological washout of lepton number especially transparent, the same approach can be applied to other processes with $\Delta L \neq 0$, $\Delta (B-L) \neq 0$ and no missing energy. Examples of such signatures are shown in Fig.~\ref{fig:decompositions2}. The leftmost diagram describes resonant pair-production into a final state with six fermions with a factorization in the same manner as Eq.~\eqref{eq:cs_lhc}. The remaining three diagrams are non-resonant variations of the topologies depicted in Fig.~\ref{fig:decompositions}. The analysis of these cases will be more involved as the total cross section cannot be easily factorized and the relation to the lepton number washout (see below) will be less transparent. However, a proportionality between the LHC cross section and the washout is still expected which allows for an analogous argumentation. In this way, the focus on a resonant process and its approximation in narrow-width are not crucial for this discussion.

\section{Washout of Lepton Number Asymmetry}
In order to connect the LHC cross section and leptogenesis, we recall the Boltzmann equations for the standard leptogenesis scenario with one heavy neutrino $N$ and neglecting flavour,
\begin{align}
\label{eq:Boltzmann1}
	H z \frac{dN_N}{dz}   = 
	- (\Gamma_D + \Gamma_S) \left( N_N - N_N^{eq} \right), \qquad
	H z \frac{dN_{L}}{dz} = 
	\epsilon \Gamma_D \left( N_N - N_N^{eq} \right) - \Gamma_W N_{L}.
\end{align}
They are generically written in terms of the heavy neutrino and $(B-L)$ number densities per co-moving volume, $N_N$ and $N_L$, respectively. The generation of a net $(B-L)$ density is driven by the decays of the heavy neutrino with the rate $\Gamma_D$. In the standard leptogenesis framework, the $CP$ asymmetry $\epsilon$ arises due to the interference of tree-level and one-loop diagrams. We define $\Gamma_D$ to contain all decays that contribute to the generation of $\epsilon$. The interaction rate $\Gamma_S$ includes all other $\Delta L = 1$ decays and scattering processes involving the heavy neutrino. The term $\Gamma_W$ contains inverse $N$ decays as well as other $\Delta L = 1,2$ scattering processes which washout the asymmetry generated in the decay. The evolution parameter $z$ is related to the temperature $T$ as $z=M_X/T$ and $N_N^{eq}$ denotes the equilibrium density of the heavy neutrinos. The temperature-dependent Hubble parameter $H$ arises due to the expansion of the Universe.

In the following, we describe how a lower limit on the washout term $\Gamma_W$ can be set based on the cross section of the previously discussed LNV process at the LHC. For this, we only take the washout $\Gamma_W$ into account, and assume $\Gamma_D = 0$ and $\Gamma_S = 0$. This implies that the lepton number density was generated at a high scale, and is washed out by the resonant process $qq \to l^\pm l^\pm q q$ alone. Other processes would only increase the washout rate and strengthen the argument. Similarly, the presence of $\Gamma_D$, e.g. due to $CP$-conserving decays of $N$, would reduce the lepton number asymmetry further. 

\begin{figure}[t]
\centering
\includegraphics[clip,width=0.45\linewidth]{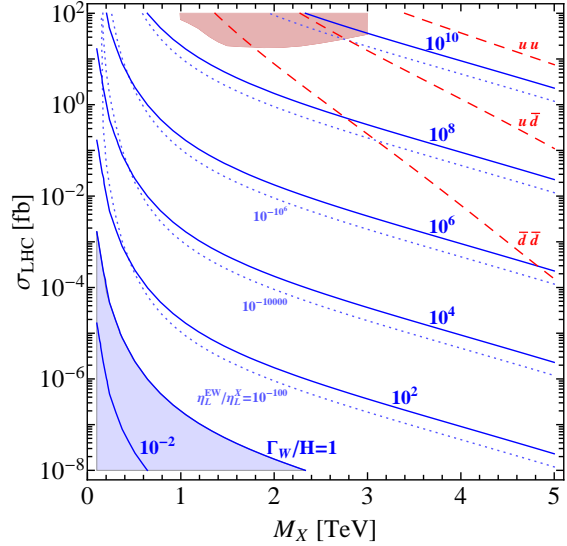}
\caption{Washout rate $\Gamma_W/H$ at $T = M_X$ as a function of $M_X$ and $\sigma_\text{LHC}$ (solid blue contours). The dotted light blue contours denote the surviving lepton asymmetry at the EW scale relative to its value at $M_X$, $\eta_L^\text{EW}/\eta_L^X$. The red dashed curves are typical cross sections of the process $pp \to l^\pm l^\pm q q$ (see text). The red shaded region at the top is excluded due to searches for resonant same sign dileptons at the LHC~\cite{CMS:2012uaa}.
}
\label{fig:GammaW_mX_sigma}
\end{figure}

The washout rate is related to the parton-level cross section $\sigma(q^2)$ (not averaged over the initial state particle quantum numbers) of the process $q q \leftrightarrow l^\pm l^\pm q q$ as
\begin{align}
\label{eq:washout_general}
  \frac{\Gamma_W}{H} \approx  
  \frac{1}{64\pi^2 H T^2}\int_0^\infty \!\! dq^2 \,\, (q^2)^{3/2} \sigma(q^2) 
	K_1\!\left( \frac{\sqrt{q^2}}{T} \right),
\end{align}
where $K_1(x)$ denotes the 1st-order modified Bessel function of the second kind. Comparing with Eq.~\eqref{eq:Boltzmann1}, a value $\Gamma_W / H \gg 1$ will lead to a rapid decrease (washout) of the $(B-L)$ asymmetry. The minimal washout rate and the LHC production cross section $\sigma_\text{LHC}$ are therefore directly related ($M_\text{P}$ is the Planck mass)~\cite{Deppisch:2013jxa},
\begin{align}
\label{eq:washout_factor_related}
  \frac{\Gamma_W}{H} &> 
  0.003\times
  \frac{M_\text{P}M_X^3}{T^4}
  \frac{K_1\!\left( M_X/T \right)}
  {f_{q_1 q_2}\!\left( M_X / \sqrt{s} \right)}
  \times(s~\sigma_\text{LHC}),
\end{align}
with the order of magnitude approximation at $T = M_X$:
\begin{align}
\label{eq:washout_factor_estimation}
  \log_{10}\frac{\Gamma_W}{H} &\gtrsim
  6.9 + 0.6\left( \frac{M_X}{\text{TeV}} - 1 \right) +
  \log_{10}\frac{\sigma_\text{LHC}}{\text{fb}}.
\end{align}
This shows that an observation of the resonant process $pp \to l^\pm l^\pm q q$ at the LHC corresponds to a very strong washout of the lepton asymmetry in the early Universe. The exact relation is shown in Fig.~\ref{fig:GammaW_mX_sigma}. 
The solid blue lines denote the corresponding washout rate as a function of $M_X$ and $\sigma_{\mathrm{LHC}}$. For any realistic cross section, which could be observable at the LHC, $\sigma_\text{LHC}\gtrsim 10^{-2}$~fb, the resulting washout is always highly effective. In order to get a feeling for the order of magnitude in specific models, the red dashed curves show examples of typical cross sections with different initial parton combinations. For these estimates, a particle $X$ with a gauge-strength total width, $\Gamma_X/M_X = g^2/(32\pi)$, with $g=0.5$ and branching ratios $\text{Br}(X\to q_1 q_2) = \text{Br}(X\to 4f) = 0.5$ has been assumed. The predictions can be compared to the limit from searches for resonant same sign dileptons at the LHC (red shaded region)~\cite{CMS:2012uaa}\footnote{CMS has reported updated results which do not affect the plot significantly, but which show a small excess that could be interpreted as a signal of $W_R$ production, although without the presence of LNV~\cite{Khachatryan:2014dka}.}. Thus, an observation of the process $pp \to l^\pm l^\pm q q$ at the LHC would necessarily result in an enormous washout of a pre-existing lepton asymmetry. This is model-independently true and purely based on the observables $M_X$ and $\sigma_\text{LHC}$ of the process. The only necessary assumption is that there is no source (re)generating the asymmetry below $M_X$.

The discussion so far neglected the role of flavour, but it is possible that leptogenesis models generate a number asymmetry in one lepton flavour only. As charged lepton flavour violation has not been observed, different flavours are not necessarily in equilibrium in the early Universe. Observing LNV at the LHC in one or two flavours is therefore not sufficient to exclude all ``flavoured'' high-scale leptogenesis scenarios. The observation of $pp \to l^\pm l^\pm q q$ in combinations involving all flavours is necessary to unambiguously falsify high-scale leptogenesis models at the LHC\footnote{Alternatively, the observation of lepton flavour violating processes such as $\tau\to\mu\gamma$ could prove that different lepton flavours are in equilibrium at certain temperatures.}. This is experimentally challenging for $\tau$ leptons.

\section{Limit on the Baryon Asymmetry}
We now additionally describe the generation of the $(B-L)$ asymmetry in the standard leptogenesis scenario with one heavy neutrino. It is parametrized by the scale $M_N$ and the $CP$ decay asymmetry $\epsilon$. As before, the washout is described by the scale $M_X$ and its LHC cross section $\sigma_{\mathrm{LHC}}$. Under these assumptions, the Boltzmann equations~\eqref{eq:Boltzmann1} (now with $\Gamma_D \neq 0$ and $\Gamma_S = 0$), can be solved and the lepton density $\eta_L = N_L/n_{\gamma}$ normalized to the photon density $n_{\gamma}$ is determined. The general behaviour is well approximated by $\eta_L \approx r_N^2 \epsilon / (z \Gamma_W / H)\exp((1-r_N)z)$ with $r_N = M_N/M_X$~\cite{Deppisch:2013jxa, Deppisch:2010fr}. 

The resulting lepton asymmetry around the critical temperature $T_c \approx 135$~GeV of the EW phase transition is converted to a baryon asymmetry, and after taking into account the evolution of the photon density until now, the lower limit Eq.~\eqref{eq:washout_factor_related} on the washout rate results in an upper limit on the final baryon asymmetry
\begin{equation}
  \label{eq:etaBapprox}
	|\eta_B| \lesssim 0.02 \times
	\frac{M_N^2}{M_X^2} \frac{T_c}{M_X} 
	\frac{|\epsilon|}
	{\Gamma_W/H}
	\exp\left(\frac{M_X-M_N}{T_c} \right),
\end{equation}
with the order of magnitude estimate
\begin{align}
  \label{eq:etaBestimation}
	\log_{10} \left|\frac{\eta_B}{\eta_B^\text{obs}}\right| \lesssim 
	2.4\, \frac{M_X}{\text{TeV}} \left( 1 - \frac{4}{3} \frac{M_N}{M_X}
\right) + \log_{10} \left[ |\epsilon|\, \left(
\frac{\sigma_\text{LHC}}{\text{fb}} \vphantom{\frac{4}{3}
\frac{M_N}{M_X}}\right)^{-1} \left( \frac{4}{3} \frac{M_N}{M_X} \right)^2
\right],
\end{align}
relative to the observed value $\eta_B^\text{obs}$. It relates the collider observables $M_X$ and $\sigma_\text{LHC}$ with the leptogenesis parameters $M_N$ and $\epsilon$. Two major conclusions can be drawn from this result: (i) As the $CP$ asymmetry $|\epsilon| \leq 1$ by definition, the observation of LNV at the LHC would exclude scenarios with $M_N \gtrsim M_X$, as it is not possible to create a large enough baryon asymmetry. This essentially reiterates the result of the previous section. (ii) For $M_N < M_X$, leptogenesis cannot be excluded model-independently as $\eta_B^\text{obs}$ can be potentially achieved for a large enough $|\epsilon|$. However, it is possible to set a stringent lower limit on $|\epsilon|$. For example, in case of a hypothetical observation of the resonant LNV process at the LHC with $M_X = 2$~TeV and $\sigma_\text{LHC} = 0.1$~fb, one has $|\epsilon| \gtrsim 10^{-3}$~\cite{Deppisch:2013jxa}.

\section{Summary and Outlook}
We have reported on the main result of our analysis \cite{Deppisch:2013jxa} which proved that the observation of LNV at the LHC would imply a large washout factor, destroy a pre-existing lepton asymmetry exponentially and render high-scale leptogenesis models ineffective. The argumentation is generally valid for any realization of high-scale thermal leptogenesis models. However, there are some caveats to be considered. Firstly, to falsify high-scale leptogenesis unambiguously, LNV in all three flavours has to be observed. Furthermore, our reasoning can only be applied to mechanisms that generate a visible lepton asymmetry (i.e. in the SM leptons) above the LHC scale. Mechanisms that (re)generate the asymmetry or that produce a net lepton number in a hidden sector which is converted to $B$ below this scale, cannot be ruled out.

Although we have concentrated on the resonant processes in Fig.~\ref{fig:decompositions}, the argumentation presented here is applicable to other LNV processes at the LHC. We want to stress that searches for LNV processes without missing energy at the LHC can have significant impact on models of leptogenesis and therefore on our understanding of baryogenesis. As the presence of LNV is closely connected to the generation mechanism of light neutrino masses, they also shed light on the nature of neutrinos. For example, the topologies in Fig.~\ref{fig:decompositions} contribute to neutrinoless double beta decay~\cite{Helo:2013ika, Deppisch:2012nb}. Thus, dedicated searches for LNV processes at colliders are promising probes of beyond the SM physics and should be further extended.

\Acknowledgements
We would like to thank Martin Hirsch for the fruitful collaboration on which this report is based. The work was supported partly by the London Centre for Terauniverse Studies (LCTS), using funding from the European Research Council via the Advanced Investigator Grant 267352. FFD would like to thank the organizers of LHCP 2014 for the pleasant conference.


\begin{thebibliography}{99}

\bibitem{Ade:2013zuv} 
  P.~A.~R.~Ade {\it et al.}  [Planck Collaboration],
  [arXiv:1303.5076 [astro-ph.CO]].

\bibitem{Gavela}
  M.~B.~Gavela, P.~Hernandez, J.~Orloff, O.~Pene and C.~Quimbay,
  Nucl.\ Phys.\ B {\bf 430} (1994) 382.
  
\bibitem{KRS} 
	V.~A.~Kuzmin, V.~A.~Rubakov, M.~E.~Shaposhnikov,
	Phys.\ Lett.\ B {\bf 155} (1985)~36.

\bibitem{Sakharov}
  A.~D.~Sakharov,
  Pisma Zh.\ Eksp.\ Teor.\ Fiz.\  {\bf 5} (1967) 32.

\bibitem{FY} 
	M. Fukugita, T. Yanagida, 
	Phys.\ Lett.\ B {\bf 174} (1986) 45.
	
\bibitem{Frere:2008ct}
  J.-M.~Frere, T.~Hambye, G.~Vertongen,
  JHEP {\bf 0901} (2009) 051;
  P.~S.~B.~Dev, C.~H.~Lee and R.~N.~Mohapatra,
  arXiv:1408.2820 [hep-ph].

\bibitem{Deppisch:2013jxa}
  F.~F.~Deppisch, J.~Harz and M.~Hirsch,
  Phys.\ Rev.\ Lett.\  {\bf 112} (2014) 221601.

\bibitem{Helo:2013ika}
  J.~C.~Helo, M.~Hirsch, H.~P\"as, S.~G.~Kovalenko,
  Phys.\ Rev.\ D {\bf 88} (2013) 073011;
  %\bibitem{Bonnet:2012kh} 
  F.~Bonnet, M.~Hirsch, T.~Ota, W.~Winter,
  JHEP {\bf 1303} (2013) 055.

\bibitem{Keung:1983uu}
	W.-Y. Keung, G.~Senjanovic,
	Phys.\ Rev.\ Lett. {\bf 50} (1983) 1427;
	%\bibitem{Das:2012ii}
  S.~P.~Das, F.~F.~Deppisch, O.~Kittel, J.~W.~F.~Valle,
  Phys.\ Rev.\ D {\bf 86} (2012) 055006.
	
\bibitem{Leike:1998wr}
  A.~Leike,
  Phys.\ Rept.\  {\bf 317} (1999) 143.

\bibitem{CMS:2012uaa}
  CMS Collaboration [CMS Collaboration],
  CMS-PAS-EXO-12-017.
  
\bibitem{Khachatryan:2014dka}
  V.~Khachatryan {\it et al.}  [CMS Collaboration],
  arXiv:1407.3683 [hep-ex];
  F.~F.~Deppisch, T.~E.~Gonzalo, S.~Patra, N.~Sahu and U.~Sarkar,
  arXiv:1407.5384 [hep-ph].

\bibitem{Deppisch:2010fr}
  F.~F.~Deppisch, A.~Pilaftsis,
  Phys.\ Rev.\ D {\bf 83} (2011) 076007.

\bibitem{Deppisch:2012nb} 
  F.~F.~Deppisch, M.~Hirsch, H.~P\"as,
  J.\ Phys.\ G {\bf 39} (2012) 124007.

\end{thebibliography}
\end{document}